\documentclass[conference]{IEEEtran}

\usepackage{graphicx}
\usepackage{amsxtra}
\usepackage{amssymb,amsmath}
\usepackage{times,amsmath,epsfig}
\usepackage{cite}
\usepackage{float}
\usepackage{subcaption}
\usepackage[noend]{algpseudocode}
\usepackage{times}
\usepackage{boxedminipage}
\usepackage{array,multirow}
\usepackage{tabularx}
\usepackage{algorithm}
\usepackage{pdfsync}
\usepackage{url}
\usepackage[nopatch=eqnum]{microtype}
\usepackage{color}
\usepackage{xpatch}
\xpretocmd{\eqref}{Eq.~}{}{}
\usepackage{acronym}

\usepackage{comment}

\acrodef{TGbf}[TGbf]{Task Group IEEE 802.11bf}

\newcommand{\remove}[1]{}
\makeatletter
\def\hlinewd#1{%
\noalign{\ifnum0=`}\fi\hrule \@height #1 %
\futurelet\reserved@a\@xhline}
\makeatother

\newcommand\blfootnote[1]{%
  \begingroup
  \renewcommand\thefootnote{}\footnote{#1}%
  \addtocounter{footnote}{-1}%
  \endgroup
}

\IEEEoverridecommandlockouts 

\begin{document}
\title{Sensing Performance of the IEEE 802.11bf Protocol and Its Impact on Data Communication}
\newcommand*\samethanks[1][\value{footnote}]{\footnotemark[#1]}

\author{\IEEEauthorblockN{
Anirudha Sahoo$^1$, Tanguy Ropitault$^{2,3}$, Steve Blandino$^{2,3}$ and Nada Golmie$^1$} 
 \IEEEauthorblockA{$^1$National Institute of Standards and Technology, Gaithersburg, Maryland, USA \\
 $^2$Associate, National Institute of Standards and Technology, Gaithersburg, Maryland, USA \\
 $^3$Prometheus Computing LLC, Bethesda, Maryland, USA \\
 Email: \{anirud, tanguy.ropitault, steve.blandino, nada\}@nist.gov}}

\maketitle
\pagenumbering{arabic}
\pagestyle{plain}


\begin{abstract}
Wi-Fi sensing has been used to detect and track movements in an environment, resulting in the emergence of several innovative applications. Wi-Fi sensing can detect movement and locate objects by analyzing variations in the Wi-Fi signal due to its interaction with moving objects. 
Until recently, Wi-Fi sensing has been primarily available through proprietary solutions, which has limited its adoption. However, the recent initiative by the IEEE to develop the IEEE 802.11bf  standard promises to make the adoption of Wi-Fi sensing widespread.
Although Wi-Fi sensing procedures in communication standards can be overhead, there is currently a lack of literature exploring the sensing performance of Wi-Fi sensing procedures specified in the IEEE 802.11bf standard and its impact on data communication.
Therefore, this paper presents a comprehensive evaluation of the sensing performance of the IEEE 802.11bf protocol and its impact on data communication in different configurations. Our findings expose the limitations
of specific configurations and pave the way
to provide guidance on 
efficient operating configurations of an IEEE 802.11bf network.

\end{abstract}


\maketitle

\section{Introduction}
\label{sec-introduction}
Using Wi-Fi sensing, changes in the Wi-Fi radio channel can be used to detect movements in the environment, enabling a wide range of applications  such as presence of humans, localization, fall detection, etc.~\cite{ma2019wifi}. Since Wi-Fi networks are widely 
deployed, the above paradigm would make it possible to make 
aforementioned diverse set of applications available to
the users and 
eliminate the need for different kinds of sensors for different applications. Although a lot of work has been
reported in the literature about Wi-Fi based sensing
~\cite{ma2019wifi, abdelnasser2015ubibreathe, li2020passive, abdelnasser2015wigest, arshad2017wi, 9989347,9477585,6212229}, lack of standardization has limited
 the proliferation of Wi-Fi sensing based applications.
Therefore, the \ac{TGbf}  started the 
development of an amendment to the IEEE 802.11 standard
in September 2020 to standardize Wi-Fi based sensing~\cite{802.11bf} which will be known as IEEE
802.11bf. The IEEE 802.11bf standard defines Wireless Local
Area Network (WLAN) sensing procedures, both in the sub-7 GHz~\cite{802.11bf, tanguy-wifi-sensing}
and above 45 GHz band~\cite{802.11bf,blandino2023ieee}
\blfootnote{U.S. Government work, not subject to U.S. Copyright.}. 

Integrating sensing with data communication in
Wi-Fi network is quite attractive since it allows for
more efficient use of spectrum and hardware. 
However, for sensing, the system generally
needs to allocate a part of its radio resources to send dedicated
sounding frames and other sensing related information,
reducing resources available for regular data communication.
Thus, sensing becomes an overhead for data communication. 
There is no study that quantifies the impact of implementing Wi-Fi sensing procedures, specified in the IEEE 802.11bf standard, on data communication.
In this work, we implement the most recent features of the IEEE 802.11bf protocol in the sub-7 GHz band by extending the \textit{IEEE 802.11ax lightsim software}~\cite{github_802.11ax} and conduct a thorough assessment of WLAN sensing performance as well as its impact on data communication. Given the novelty of a Wi-Fi network performing sensing procedures, we also introduce new performance metrics, which 
are designed to quantify both the overhead introduced by the sensing procedures and the failure rate of the sensing operations.

In the IEEE 802.11bf sensing
protocol for sub-7 GHz, the actual sensing measurements are done during, what are called, \textit{sensing measurement exchanges} (SMEs), and those SMEs account for most of the sensing overhead. So, in this study, we focus on the SME part of the protocol. To perform an SME, the initiator of sensing, which could be the Access Point (AP) or a Wi-Fi Station (STA), has
to get access to the channel and obtain a Transmission Opportunity (TxOP). It may use Enhanced Distributed Channel Access (EDCA) or 
Point coordination function (PCF) Interframe Space (PIFS) to get a TxOP. The IEEE 802.11bf protocol also defines a periodically
occurring sensing window, within which the SMEs have to be 
performed. The duration and periods of sensing windows are
configurable. We examine
both the EDCA and PIFS based access methods
at different sensing loads in the system when the AP is the initiator of sensing.
We present extensive simulation results at different system
configurations corresponding to different sensing loads.
Our simulation results have uncovered significant findings that will facilitate the efficient configurations of future IEEE 802.11bf systems operating in the sub-7 GHz band.

The main contributions of this work are as follows.
\begin{itemize}
    \item To the best of our knowledge, this is the first work to present an extensive simulation of the IEEE 802.11bf protocol.
    \item This work provides quantitative insights into the sensing performance of the IEEE 802.11bf protocol and its impact on data communication in terms of the defined performance metrics.
    \item Our simulation exposes the limitations of WLAN sensing using EDCA based access when sensing reports need to be sent from sensing STAs to the AP.
    \item The results presented in this work provide guidance and insights into efficient operating configurations of an IEEE 802.11bf network.
\end{itemize}

\section{Related Work}
\label{sec-relatedwork}
There has been a lot of research work on Wi-Fi sensing reported in the literature. In~\cite{li2020passive}, the authors present a passive
Wi-Fi radar system for human sensing by exploiting high data rate
OFDM signals and periodic Wi-Fi beacon signals. Change in 
Received Signal Strength (RSS) in a commercial off-the-shelf (COTS) Wi-Fi device
held on a person's chest is used to design a respiratory monitoring system in~\cite{abdelnasser2015ubibreathe}. Changes in Wi-Fi
signal strength have been studied to detect hand gestures around a
user's mobile device in~\cite{abdelnasser2015wigest}. In~\cite{9989347}, the authors have implemented an end-to-end system to monitor human
respiratory motion using Wi-Fi Channel State Information (CSI). They propose a deep
learning based processing algorithm called \textit{BreatheSmart} that analyzes the changes in amplitude and phase of CSI data to detect respiratory motion.
In~\cite{9477585}, a four antenna passive bistatic indoor radar configuration is set up
using IEEE 802.11ax Wi-Fi system to track  multi-target human
based on range, doppler and
angle-of-arrival measurements. A prototype of Wi-Fi based passive radar system
for localization and tracking of moving targets using range, doppler
and direction of arrival is presented in~\cite{6212229}. The above mentioned research
works are focused on methodologies or algorithms for
the concerned applications, but do not deal
with estimating the Wi-Fi sensing related
overhead of the system.
A fairly
comprehensive survey of Wi-Fi sensing with CSI is presented in~\cite{ma2019wifi}.

To standardize Wi-Fi sensing process, the \ac{TGbf} is developing a standard which will be known as IEEE 802.11bf~\cite{802.11bf}. This standard defines the mechanisms
and protocols to provide channel state information in the sub-7
GHz band and radar based information (e.g., range, doppler, beam
azimuth) above the 45~GHz band. Since Wi-Fi sensing protocol is an overhead to the Wi-Fi data communication, it is important to
study the performance of Wi-Fi sensing and
its impact on data communication in different configurations. To the best of our knowledge, there is no such study available
in the literature. 



\section{Overview of IEEE 802.11bf Sensing Procedure}
\label{sec-overview}
\begin{figure}[!t]
\centering
\includegraphics[scale=0.4]
{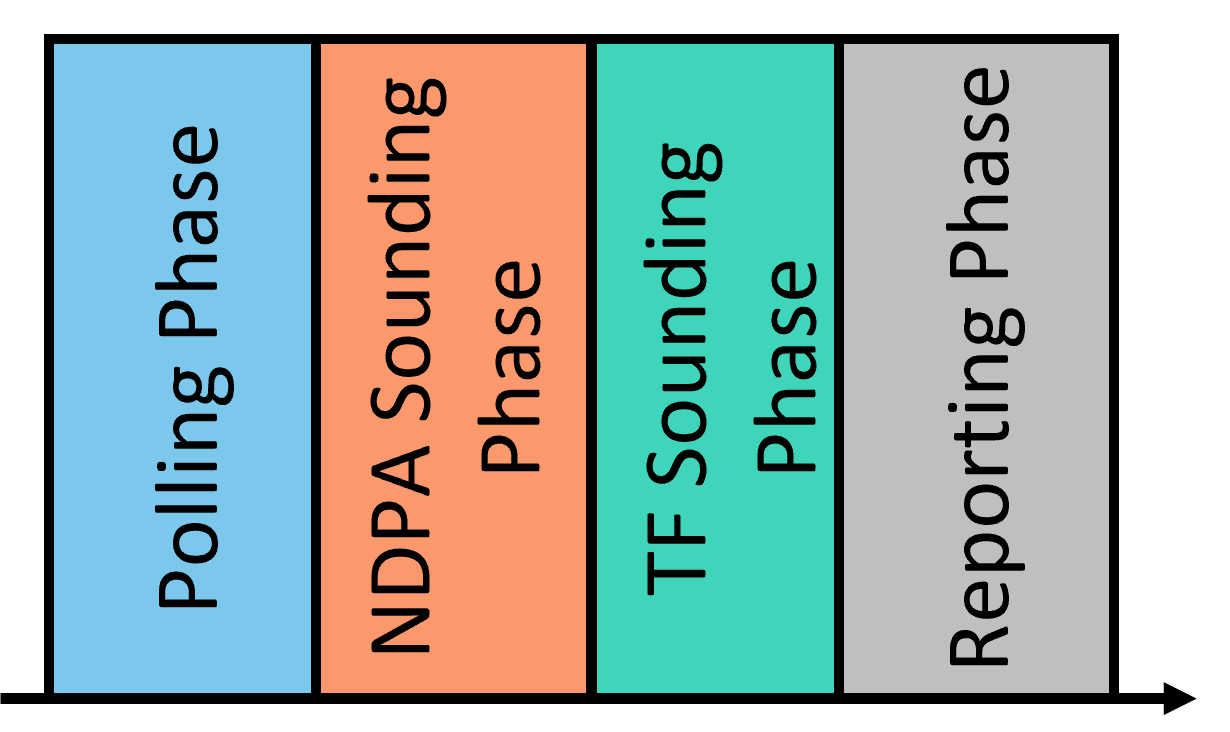}
\caption{Different phases in a TB Sensing Measurement Exchange}
\label{fig:tb_meas_phases}
\end{figure}

An IEEE 802.11bf capable STA and AP
exchange their sensing capabilities during the association process.  WLAN sensing in the sub-7 GHz band, referred to as
\textit{Sensing Procedure}, starts out with the  establishment of a
\textit{sensing measurement session} between a \textit{sensing initiator}
and a \textit{sensing responder} at which time the operating
parameters of the session are determined. Examples of
operating parameters include bandwidth, the role of the
STAs (sensing transmitter or sensing receiver), timer values etc. The actual
sensing measurements are performed in SMEs. SMEs
can be trigger based (TB) or non-trigger based (NTB). 
Since TB SME is
envisioned to be the most common deployment scenario, we focus
on TB SME in this study.
In a TB SME, an AP is the
sensing initiator and one or more non-AP STAs are the sensing
responders. An AP starts a TB SME by obtaining a TxOP after a
\textit{sensing availability window} (SAW) period starts. A SAW has two parameters:
\textit{SAW duration} and \textit{SAW period}.
SAW durations occur periodically and their periods are
determined by the SAW period parameter. An AP and the STAs may participate in SMEs only in a SAW duration.
A TB SME can have up
to four phases as shown in Fig.~\ref{fig:tb_meas_phases}.
In the polling phase, an AP (the sensing initiator) sends a Sensing Polling Trigger
frame to the sensing responder STAs inviting them to participate in the SME.  In the Null Data Packet Announcement (NDPA) Sounding phase
the AP is the sensing transmitter and one or more STAs are
the sensing receivers. The AP sends an NDPA frame followed by a Null Data Packet (NDP) frame to the receiver STAs. The STAs measure the channel state using the received NDP
frame. In the Trigger Frame (TF) Sounding phase, the AP
acts as the sensing receiver and the STAs as sensing transmitters. The AP sends a TF to the sensing receiver
STAs, which then send NDP frames (which are multiplexed in
the spatial domain) to the AP. The AP measures the channel
state using the received NDP frames. The reporting phase is present, if sensing report (mainly consisting of CSI) is required to be sent from the STAs to the AP. Orthogonal Frequency Division Multiple Access (OFDMA) mechanism is used for reporting, for which the AP allocates
Resource Units (RUs) to the STAs. Only the NDPA sounding phase may necessitate a Reporting phase, if reporting was enabled as part of operational parameters during the sensing measurement session setup. For more
details on TB SME, please refer to~\cite{802.11bf}.

\begin{figure}[htb]
    \centering
    \begin{subfigure}{0.25\textwidth}
        \centering
        \includegraphics[width=\textwidth]{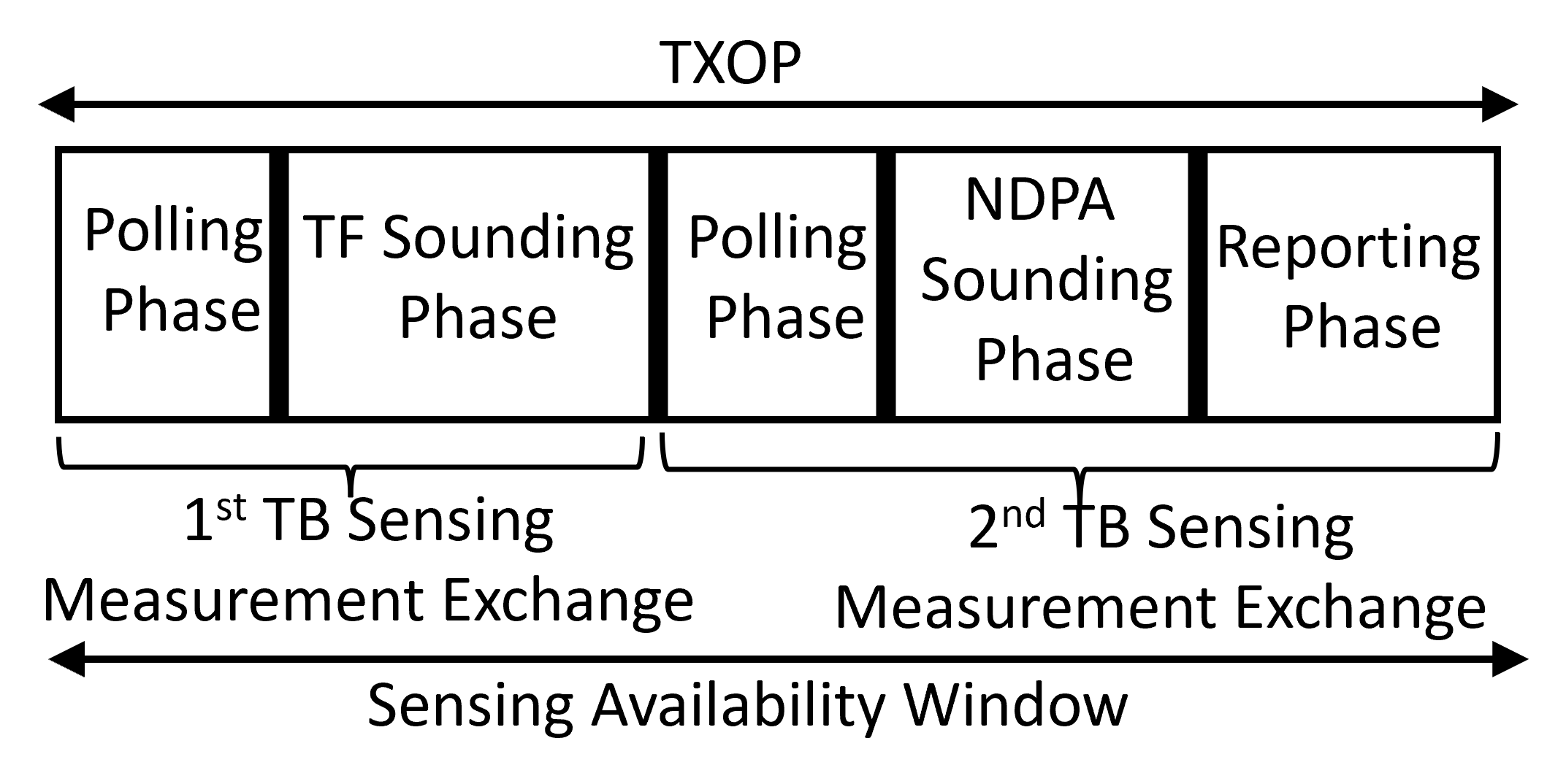}
        \captionsetup{width=0.8\linewidth}
        \caption{SAW with two SMEs within a single TxOP}
        \label{fig:saw_one_txop}
    \end{subfigure}%
    \qquad
    \begin{subfigure}{0.4\textwidth}
        \centering 
        \includegraphics[width=\textwidth]{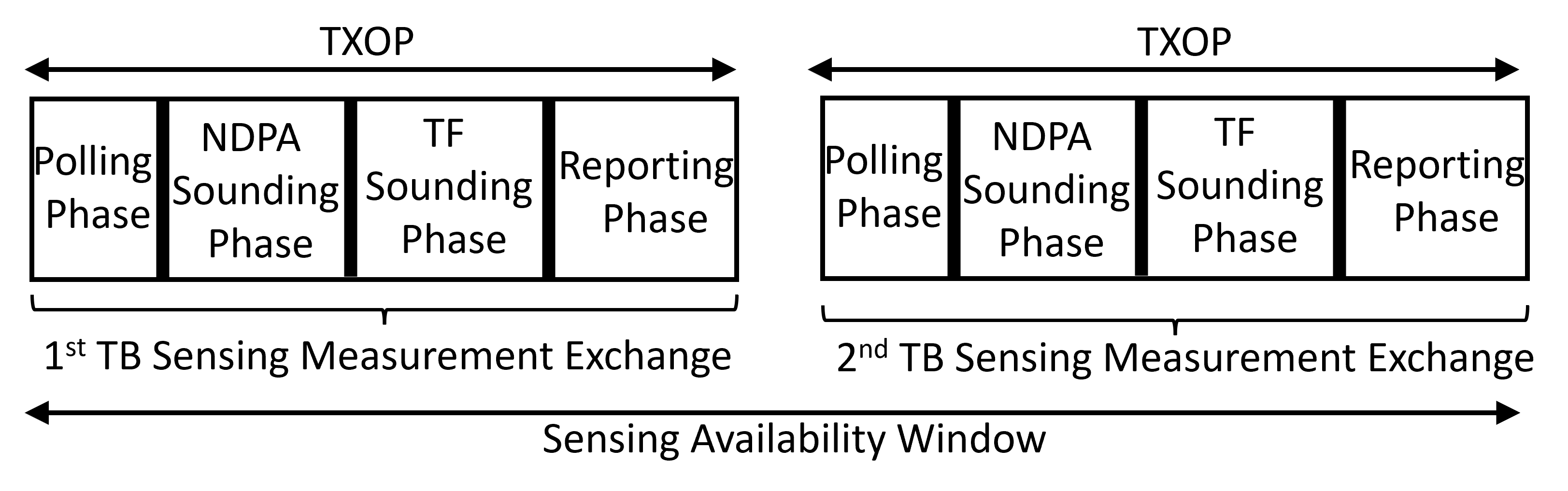}
        \caption{SAW Spanning across two TxOPs, each with an SME}
        \label{fig:saw_two_txop}
    \end{subfigure}
    \caption{Illustration of SAW, TxOP and SMEs}
    \label{fig:nsta_pso}
\end{figure}

As mentioned earlier, an AP performs TB SMEs within a TxOP, after obtaining the TxOP within a SAW. A SAW may contain a single TxOP (Fig.~\ref{fig:saw_one_txop}) or more than one TxOP (Fig.~\ref{fig:saw_two_txop}).
An AP may obtain the TxOP by EDCA or PIFS mechanism. We refer
to them as \textit{EDCA access} and \textit{PIFS access} 
respectively. If it
uses EDCA access, then due to contention, the actual SAW duration 
available for SMEs sometimes may be shorter than the 
configured SAW duration. But
when PIFS access is used, AP gets priority access to the channel
and hence, gets almost the entire SAW duration for sensing.

\subsection{Overhead Calculation}
\label{sec:ovrhd}
The Sensing Procedure is an overhead for data communication. The majority of
the overhead is incurred in the SME part of the protocol. 
So, in this study, we concentrate on the SME part of the protocol.
In an SME, the NDPA sounding phase and Reporting phase account for
most of the overhead. Hence,
our overhead calculation involves only those two phases. Note that in an NDPA sounding phase, the AP acts as a sensing transmitter, and one or more STAs act as sensing receivers. For the NDPA
sounding phase, we take the number of bytes in NDPA and NDP frame
structure as overhead~\cite{802.11bf}. The NDPA frame structure is 
presented in Fig.~9.58 in~\cite{802.11standard_2020}
and the STA Info field format used in the NDPA frame is shown in Fig.~9-61da in~\cite{802.11az}. The NDP format shown in Fig.~27.46a in~\cite{802.11az} is used for the computation of NDP overhead. For the reporting overhead, we use the CSI size computation used in Equation~(9-5f) in~\cite{802.11bf}:

\begin{equation}
    \begin{split}
CSI \, size = & \lceil 1.5 \times N_{tx} \times  N_{rx} \rceil + \frac{N_{tx} \times N_{rx} \times N_b \times N_{sc}}{4} \\
      & + 2 \times N_{rx}, 
    \end{split}
\end{equation}
where
$N_{tx}$ is the number of transmit antennas, $N_{rx}$ is the number of receive antennas, $N_b$ is the number of bits used for quantization of each CSI value, $N_{sc}$ is the number of subcarriers reported in CSI. The bytes transmitted as part of NDPA, NDP, and reporting frame will be referred to by a general term called \textit{sensing information bytes} throughout this paper.


\section{Simulation Experiments}
\subsection{Simulation Setup}
In our simulation setup, in terms of network topology, we assume that there is one
AP and a variable number of STAs associated with the AP.
Our simulation  assumes
that all messages are received correctly by the receiver, i.e., 
there is no message error due to interference.

We assume that each sensing application runs on every STA in the 
network and that the  AP requires CSI report from every
STA in the network for a given sensing application. Due to 
resource limitations, if a complete report cannot be sent
from a STA, the STA still sends a partial report. Although this will not happen in practice, we resort
to this method to highlight the sensing overhead and the missed
sensing that such cases lead to. When there is no sensing
activity in the network, the STAs send data traffic using EDCA with full bandwidth. We assume that each STA always has at least a TxOP worth of data to send. The TxOP duration was set to its 
maximum value of 5.484\,ms~\cite{daldoul2020performance}. The AP only participates in sensing and does not 
send any data traffic.


Since the SME part of the Sensing
Procedure incurs the most overhead, this has the most
impact on data communication. Hence, it is the focus
of our simulation. As mentioned in Section~\ref{sec:ovrhd},
we compute overhead based on the NDPA sounding and Reporting phase of an SME. During the Reporting phase, the AP allocates one RU
(RU sizes given in Table~\ref{tab:subcarrier}) to
each STA, which is used by each STA to send sensing reports using OFDMA.
We have developed our simulator by extending the   
\textit{lightsim} software, which was used in the simulation study reported in~\cite{daldoul2020performance}, with IEEE 802.11bf features~\cite{github_802.11ax}.

\begin{table}[htb]
\centering
\caption{Subcarrier Allocation vs. Number of STAs}
\label{tab:subcarrier}
\begin{tabular}{|p{0.4\columnwidth}|p{0.5\columnwidth}|}
\hline
\textbf{Number of STA}   & \textbf{Subcarriers per STA (size of RU allocation per STA)} \\ \hline
  1  &        996 \\ \hline
  2  &        484 \\ \hline  
  [3 - 4]  &        242 \\ \hline
  [5 - 9]  &        106 \\ \hline
  [10 - 16]  &        52 \\ \hline
\end{tabular}
\end{table}

\subsection{Performance Metrics}
Since this is the first work that evaluates the 
performance of the IEEE 802.11bf protocol, 
no pre-existing performance metrics are available for this study. Hence,
we define the following performance metrics.
\begin{itemize}
    \item Percentage Sensing Overhead (PSO): It is the percentage of total simulation duration spent on exchanging sensing information bytes. 
    
    \item Percentage Sensing Missed (PSM): In every SAW period, sensing is considered to be complete, if all the sensing information bytes for all the applications are able to be sent in the SAW duration. If no sensing messages were sent (completely missed) or only a part of sensing messages were sent (partially missed), then we consider those cases as
    \textit{sensing missed}. So, the percentage of the number of SAW periods in which sensing is missed is defined as Percentage Sensing Missed (PSM).
    
     \item Data Throughput: This is the total number of data bits sent by all the STAs divided by the simulation time.

     \item Percent Available Window Duration (PAWD): This is defined as the percentage of SAW duration actually available for sensing related tasks. Note that 
     when EDCA access is used, some part of the SAW duration
     may be lost due to contention. The PAWD would fall below 100\,\% in such situations.
\end{itemize}

\begin{figure}[htb]
    \centering
    \begin{subfigure}{0.35\textwidth}
        \centering
        \includegraphics[width=\textwidth]{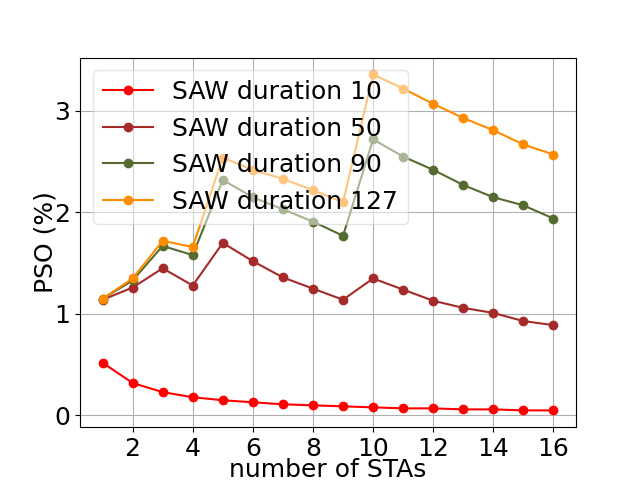}
        \caption{EDCA Access}
        \label{fig:edca_nsta_pso}
    \end{subfigure}%
    \qquad
    \begin{subfigure}{0.35\textwidth}
        \centering 
        \includegraphics[width=\textwidth]{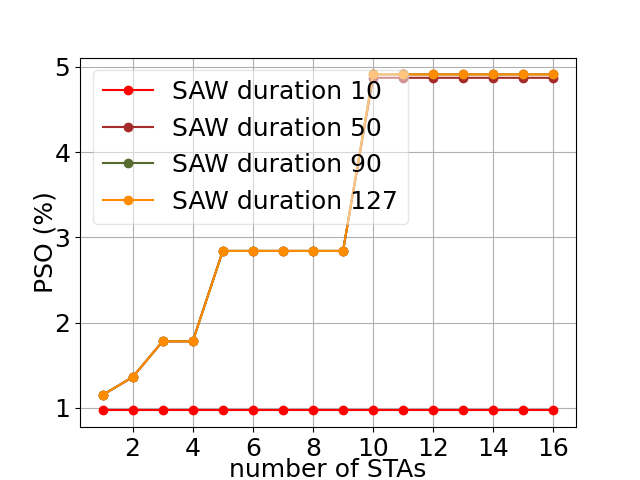}
        \caption{PIFS Access}
        \label{fig:pifs_nsta_pso}
    \end{subfigure}
    \caption{Sensing Overhead vs number of STAs (Configuration 1)}
    \label{fig:nsta_pso}
\end{figure}

\begin{figure}[htb]
    \centering
    \begin{subfigure}{0.35\textwidth}
       \centering 
        \includegraphics[width=\textwidth]{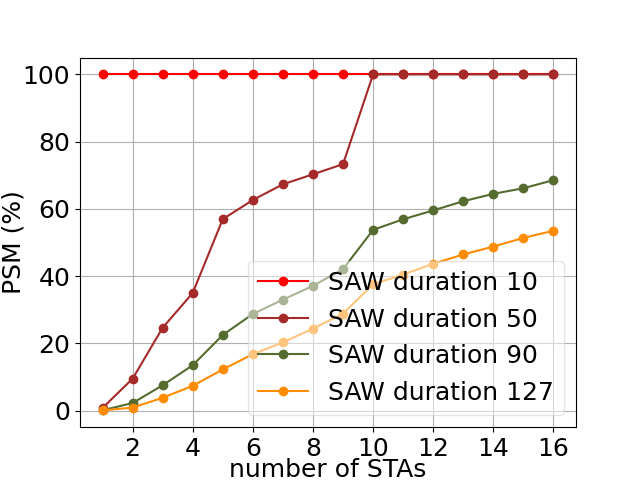}%
        \caption{EDCA Access}
        \label{fig:edca_nsta_psm}%
    \end{subfigure}
    \qquad
    \begin{subfigure}{0.35\textwidth}%
        \centering
        \includegraphics[width=\textwidth]{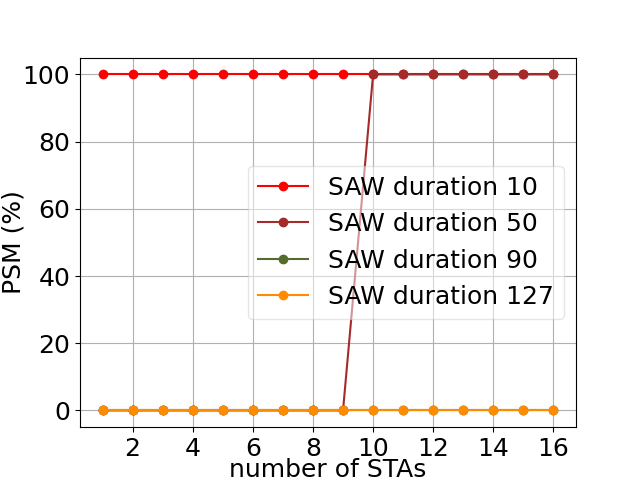}%
        \caption{PIFS Access}
        \label{fig:pifs_nsta_psm}
    \end{subfigure}
    \caption{Missed Sensing vs. number of STAs (Configuration 1)}
    \label{fig:nsta_psm}
\end{figure}

\begin{figure}[htb]
    \centering
    \begin{subfigure}{0.35\textwidth}%
      \centering
        \includegraphics[width=\textwidth]{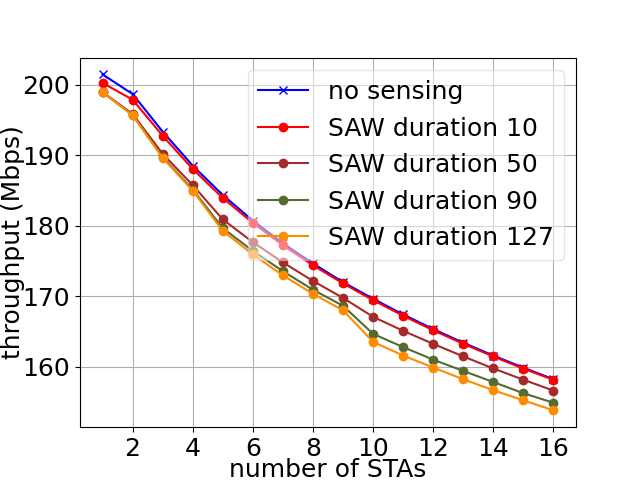}%
        \caption{EDCA Access}
        \label{fig:edca_nsta_thrpt}%
    \end{subfigure}
    \qquad
    \begin{subfigure}{0.35\textwidth}%
        \centering
        \includegraphics[width=\textwidth]{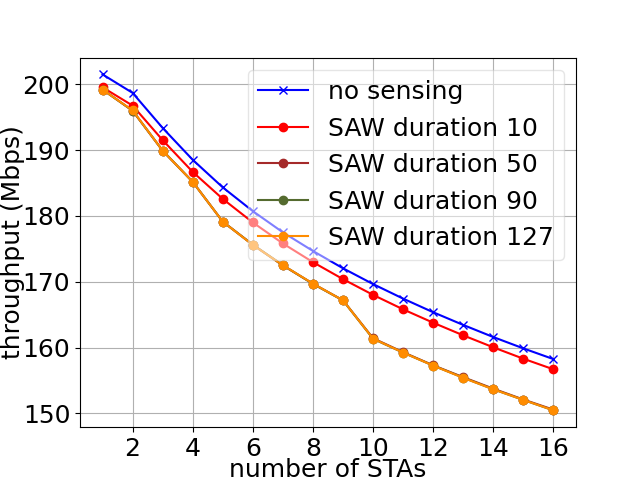}%
        \caption{PIFS Access}
        \label{fig:pifs_nsta_thrpt}%
    \end{subfigure}
    \caption{Throughput vs. number of STAs (Configuration 1)}
    \label{fig:nsta_thrpt}
\end{figure}

\begin{table}[htb]
\centering
\caption{Simulation Parameters Common to all Configurations}
\label{tab:sim_params}
\begin{tabular}{|p{0.4\columnwidth}|p{0.5\columnwidth}|}
\hline
\textbf{Parameter}   & \textbf{Value} \\ \hline
  Sensing availability window period &        1 (=100 TU = 102.4 ms) \\ \hline    

  TxOP duration & 5.484 ms \\ \hline
  
  Number of antennas in the AP &  8 \\ \hline
  
  Number of antennas in each STA   &   2 \\ \hline
  
  AP Bandwidth          & 80 MHz\\  \hline
  STA Bandwidth          & 80 MHz\\  \hline
  maximum number of subcarriers  & 996 \\ \hline
  subcarrier grouping ($N_g$) & 4 \\ \hline
  Number of subcarriers reported in CSI ($N_{sc}$) & 250 \\ \hline
  Number of bits used for quantization of each CSI value ($N_b$) & 8 \\ \hline
  EDCA transmission in a TxOP & Payload  = 10 ethernet packets of size 1500 bytes each in an A-MPDU\\ \hline
  MCS  & 6 \\ \hline
  Simulation duration & 10000 seconds \\ \hline
\end{tabular}
\end{table}

\subsection{Simulation Experiment Design}
Three parameters decide the sensing load in an IEEE 802.11bf network: (i) the number of sensing STAs, (ii) the number of sensing applications, and (iii) the number of
transmit and receive antennae involved in sensing.
Hence, for this study, we increased the sensing load in the system by increasing the value of one of those parameters while keeping the values of the other two constant.
This led us to run our experiments in three
configurations. However, due to space limitation, we are not able to present the results of the third configuration in which sensing  load was increased by increasing the number of
transmit and receive antennae. The two configurations presented in this paper are described below.
\begin{itemize}
    \item Configuration 1: In this configuration, 
    sensing load is increased by increasing the number of STAs ($\mathit{nSTA}$s) involved in sensing at different SAW durations. The number of applications is fixed at 4, and the sensing transmitter and receiver antenna configuration is set to 2x2. Note that sensing transmitter and receiver antenna configuration 2x2 implies that for each STA, the AP (sensing transmitter) uses two of its eight antennae, and each STA (sensing receiver) uses all of its two antennae. Thus, the AP can engage with up to four STAs in an SME for sensing.
    
    \item Configuration 2: Sensing load, in this configuration,  is increased by increasing the number of sensing applications ($\mathit{numApp}$) in the system at different $\mathit{nSTA}$ values. The SAW duration is fixed at 127 (Note: SAW duration 1 = 100\,$\mu$s), which corresponds to its maximum possible value of 12.7\,ms. The sensing transmitter and receiver antenna configuration is set to 2x2.
\end{itemize}

Simulation parameters common to all configurations
are shown in Table~\ref{tab:sim_params}. 

\subsection{Experiment Results}
\label{sec:results}

\subsubsection{Configuration 1}


Fig.~\ref{fig:edca_nsta_pso} shows how PSO
changes as $\mathit{nSTA}$ increases with
EDCA access. Generally, PSO decreases as $\mathit{nSTA}$ increases, because there is more contention for getting TxOP for sensing, and hence, less duration is available for sensing. However, PSO increases from $\mathit{nSTA}$ = 4 to 5 and from $\mathit{nSTA}$ = 9 to 10 for SAW duration $>$ 10. At these $\mathit{nSTA}$ transition points, the size of an RU assigned to each STA goes down (see Table~\ref{tab:subcarrier}). Hence, more time is needed to send a given number of sensing information bytes, thereby increasing the overhead.
SAW duration 10 (1 ms) is very short relative to the SAW period of 102.4 ms. Hence, the overhead is very low in this case, and at
high $\mathit{nSTA}$, due to high contention, PSO goes down to almost zero.

As seen in Fig.~\ref{fig:pifs_nsta_pso}, with PIFS access, PSO remains unchanged when RU size per STA does not change. 
Unlike EDCA access, there is no variability in actual SAW
duration available for sensing since no contention is involved. 
Report size per STA does not change since the number of applications is constant in this configuration. Hence, PSO remains constant in the intervals where RU size per STA does not change. 
But when RU size per STA decreases (e.g., 
from $\mathit{nSTA}$ = 9 to 10), the duration to send sensing report goes up, and hence, PSO goes up.
SAW duration = 10 is too short, which limits the number of sensing information bytes sent to a constant value across the $\mathit{nSTA}$ values, and hence, PSO does not change. Note that PSO for SAW durations 90 and 127 are identical all throughout. For these two SAW durations
PSM is 0\,\% all throughout (see Fig.~\ref{fig:pifs_nsta_psm}). Hence, the amount
of sensing information bytes sent is the same for the two SAW durations. For SAW duration
50, PSO is identical to those of SAW duration 90 and 127 until $\mathit{nSTA}$=9, since PSM is 0\,\% until that point. But after that, PSM goes up to  100\,\%. But these
missed sensing are due to partial missed sensing, and PSO beyond $\mathit{nSTA}$=9 is just 0.04\,\% lower than those of SAW durations 90 and 127. Hence, its PSO looks almost identical to them after $\mathit{nSTA}$=9. This indicates that SAW duration 50 fell slightly short of the duration needed to send all the sensing information bytes.

 With EDCA access, as $\mathit{nSTA}$ increases, PSM increases (see Fig.~\ref{fig:edca_nsta_psm}). Due to more contention, the actual SAW duration available for sensing
decreases, hence, more sensing is missed. SAW duration 10 is too short for EDCA, so $100\,\%$  sensing is always missed.
Except for $\mathit{nSTA}$ = 1 none of the configurations can give $0\,\%$  PSM, which is important for sensing
application performance.
For SAW duration = 10, even though PSM is $100\,\%$, there is overhead, which is due to partially missed sensing.

SAW duration 10 is too short even for PIFS access. Hence, $100\,\%$  sensing is missed (see Fig.~\ref{fig:pifs_nsta_psm}). But SAW duration 90 and 127 give $0\,\%$ throughout. Beyond $\mathit{nSTA}$ = 9, SAW duration 50 is not long enough, to send all the sensing information bytes due to a decrease in RU size per STA.

With EDCA access, from Fig.~\ref{fig:edca_nsta_thrpt}, it can be seen that the throughput goes down when sensing is on (compared to no sensing).  As $\mathit{nSTA}$ increases, the throughput decreases due to more contention and collisions. Also, as SAW duration increases, the throughput decreases because more time is used for sensing. For SAW duration 10 and $\mathit{nSTA}$ $\geq$ 3, the throughput almost equals that of no sensing case because the actual available sensing duration becomes very short due to higher TxOP contention.

As shown in Fig.~\ref{fig:pifs_nsta_thrpt}, with PIFS access, the throughput is lower than the respective EDCA cases due to higher sensing overhead (and lower missed sensing). The throughput of SAW durations 50, 90, and 127 are almost equal since sensing overheads for these cases are nearly equal.

\begin{figure}[htb]
    \centering
    \begin{subfigure}{0.35\textwidth}
    \centering
        \includegraphics[width=\textwidth]{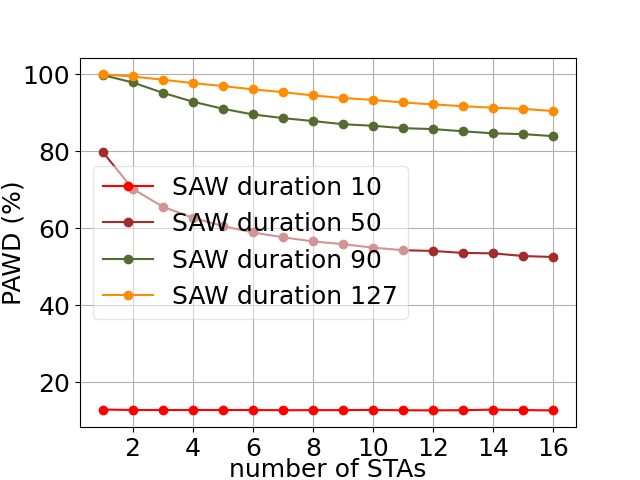}%
        \caption{Configuration 1}
        \label{fig:config1_pawd}
    \end{subfigure}
    \qquad
    \begin{subfigure}{0.35\textwidth}
    \centering
        \includegraphics[width=\textwidth]{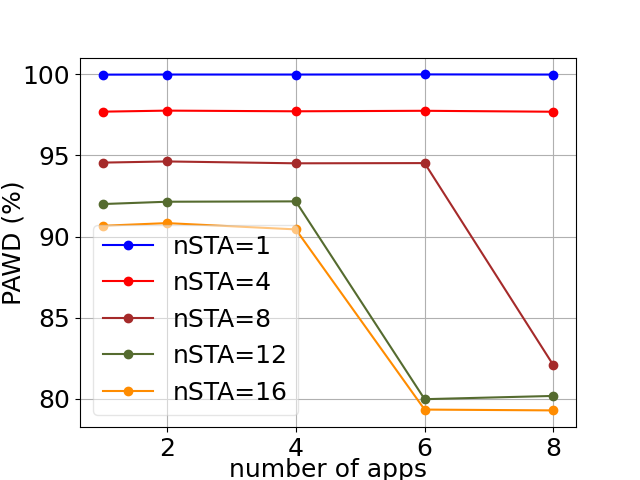}%
        \caption{Configuration 2}
        \label{fig:config2_pawd}
    \end{subfigure}
    \caption{Available Window Duration vs. number of STAs (EDCA Access)}
    \label{fig:pawd}
\end{figure}

With EDCA access, PAWD generally decreases as $\mathit{nSTA}$ increases due to an increase in contention (see Fig.~\ref{fig:config1_pawd}). As expected, the higher the SAW duration, the higher the PAWD.
PAWD is always found to be $100\,\%$ or very close to $100\,\%$ for PIFS access (not shown in a graph).

\subsubsection{Configuration 2}

\begin{figure}[htb]
    \centering
    \begin{subfigure}{0.35\textwidth}
        \centering
        \includegraphics[width=\textwidth]{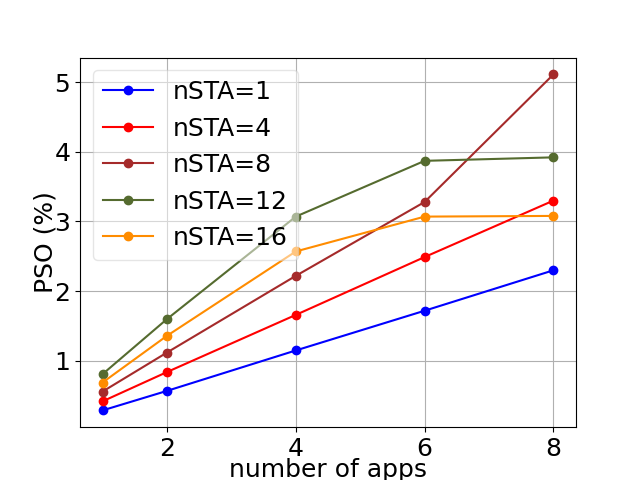}
        \caption{EDCA Access}
        \label{fig:edca_numapp_pso}%
    \end{subfigure}
    \qquad
    \begin{subfigure}{0.35\textwidth}
        \centering
        \includegraphics[width=\textwidth]{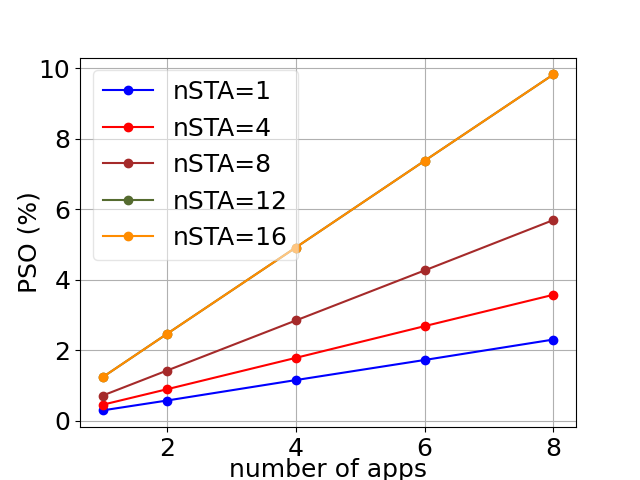}
        \caption{PIFS Access}
        \label{fig:pifs_numapp_pso}%
    \end{subfigure}
    \caption{Sensing Overhead vs number of Apps (Configuration 2)}
    \label{fig:numapp_pso}
\end{figure}

\begin{figure}[htb]
    \centering
    \begin{subfigure}{0.35\textwidth}
        \centering
        \includegraphics[width=\textwidth]{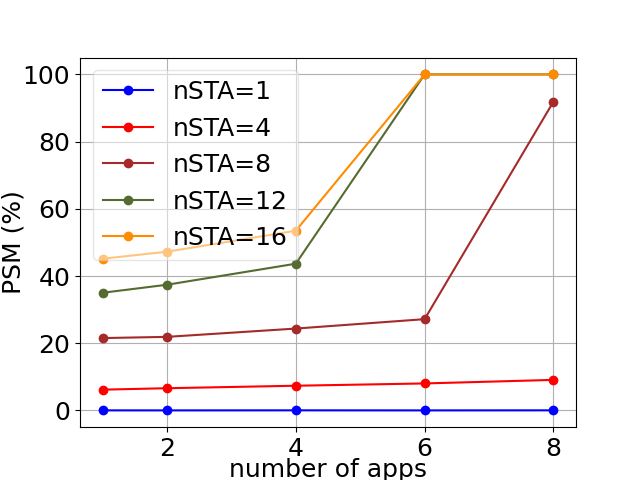}%
        \caption{EDCA Access}
        \label{fig:edca_numapp_psm}%
    \end{subfigure}
    \qquad
    \begin{subfigure}{0.35\textwidth}
        \centering
        \includegraphics[width=\textwidth]{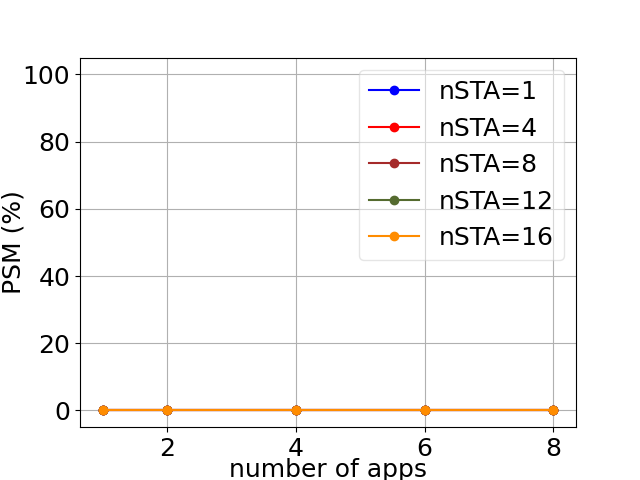}%
        \caption{PIFS Access}
        \label{fig:pifs_numapp_psm}%
    \end{subfigure}
    \caption{Missed Sensing vs. number of Apps (Configuration 2)}
    \label{fig:numapp_psm}
\end{figure}

\begin{figure}[htb]
    \centering
    \begin{subfigure}{0.35\textwidth}
       \centering
        \includegraphics[width=\textwidth]{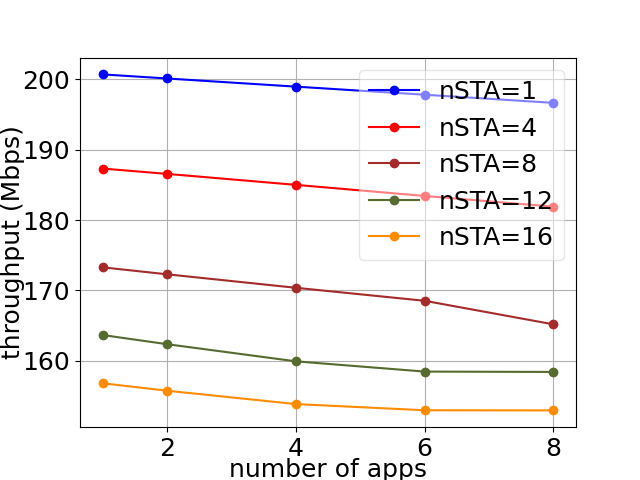}%
        \caption{EDCA Access}
        \label{fig:edca_numapp_thrpt}%
    \end{subfigure}
    \qquad
    \begin{subfigure}{0.35\textwidth}
        \centering
        \includegraphics[width=\textwidth]{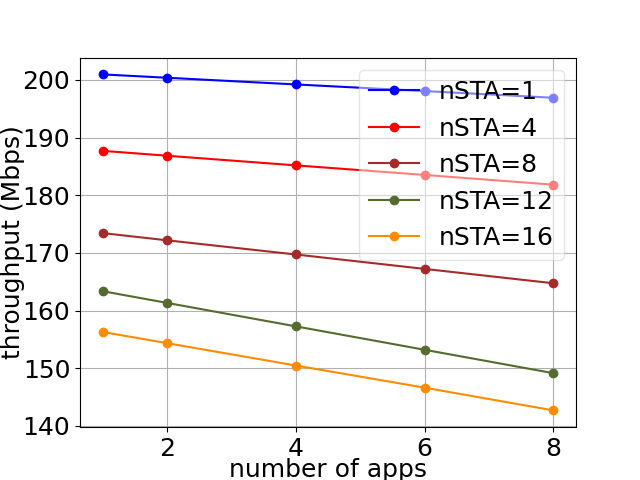}%
        \caption{PIFS Access}
        \label{fig:pifs_numapp_thrpt}%
    \end{subfigure}
    \caption{Throughput vs. number of Apps (Configuration 2)}
    \label{fig:numapp_thrpt}
\end{figure}

With EDCA access, PSO generally increases as $\mathit{numApp}$ increases (Fig.~\ref{fig:edca_numapp_pso}). At high $\mathit{numApp}$ (e.g., 6 and 8), for $\mathit{nSTA}$ = 12 and $\mathit{nSTA}$ = 16, PSO remains flat, because the number sensing information bytes that can be sent in the SAW duration is limited by the RU size allocated to the STA. This can also be explained through  PSM graph in Fig.~\ref{fig:edca_numapp_psm} where between $\mathit{numApp}$ = 6 and 8, PSM becomes $100\,\%$ for $\mathit{nSTA}$ = 12 and 16. We notice that the overhead of $\mathit{nSTA}$ = 12 is more than $\mathit{nSTA}$=16, which is counterintuitive. With $\mathit{nSTA}$ = 16, there is less duration available for sensing due to more contention. Hence, $\mathit{nSTA}$ = 12 gets more sensing opportunities and incurs higher overhead. We also notice that for $\mathit{nSTA}$ = 8, overhead goes beyond $\mathit{nSTA}$ = 12 and 16 at $\mathit{numApp}$ = 8. $\mathit{nSTA}$ = 8 has a larger RU size than $\mathit{nSTA}$ = 12 and 16, and at $\mathit{numApp}$=8 there are more sensing information bytes to be sent than at lower $\mathit{numApp}$ values. Hence, more sensing information bytes could be sent at $\mathit{nSTA}$ = 8 than at $\mathit{nSTA}$ = 12 and 16.

In the case of PIFS access, overhead consistently increases as $\mathit{numApp}$ increases and $\mathit{nSTA}$ increases (see Fig.~\ref{fig:pifs_numapp_pso}). This can be explained by observing PSM (see Fig.~\ref{fig:pifs_numapp_psm}), where there is no sensing missed. Hence, overhead increases with an increase in $\mathit{numApp}$ and also with an increase in $\mathit{nSTA}$. 

Fig.~\ref{fig:edca_numapp_psm} shows that with EDCA access, PSM increases as $\mathit{numApp}$ increases. At some $\mathit{nSTA}$ values, the jump is more drastic at certain $\mathit{numApp}$. For example, for $\mathit{nSTA}$ = 12 and 16, as $\mathit{numApp}$ increases from 4 to 6, the report size increases such that with the allocated RUs, 
the full report cannot be sent even for one application. 
Hence, PSM increases drastically to $100\,\%$.  For $\mathit{nSTA}$ = 1 and 4, the $\mathit{numApp}$ increase does not affect PSM due to low report size.

In the case of PIFS access (see Fig.~\ref{fig:pifs_numapp_psm}), there is no missed sensing in any configuration since PIFS gives priority access to the channel and the SAW duration 127 is long enough to
send all the sensing information bytes.

With EDCA access (see Fig.~\ref{fig:edca_numapp_thrpt}), for a given $\mathit{nSTA}$, the throughput goes down 
slowly as $\mathit{numApp}$ increases since it is only affected by the report size increase. But for a given $\mathit{numApp}$, as $\mathit{nSTA}$ increases, the throughput drops much more due to higher contention and collisions as well as due to report size increase. For $\mathit{nSTA}$ = 12 and 16, as $\mathit{numApp}$ increases from 6 to 8, the throughput remains flat because the PSO, in this case, does not change (see Fig.~\ref{fig:edca_numapp_pso}).

From Fig.~\ref{fig:pifs_numapp_thrpt}, we observe that with PIFS access, the throughput decrease is steeper than EDCA access as $\mathit{numApp}$ increases, since PIFS access incurs $0\,\%$ PSM and higher sensing overhead than EDCA access. 


Fig.~\ref{fig:config2_pawd} shows the PAWD performance for EDCA access. Since the SAW duration is 12.7 ms and TxOP is 5.484 ms, 
sensing can have up to three TxOPs. When $\mathit{nSTA}$ is small (1 and 4), then increasing $\mathit{numApp}$ does not change the duration and the number of TxOPs required to complete sensing. Hence, PAWD remains almost constant. But at large $\mathit{nSTA}$ and large $\mathit{numApp}$, (e.g., $\mathit{nSTA}$ = 12 and $\mathit{numApp}$ = 6), it requires more TxOPs to finish sensing operation. Since each TxOP is subject to contention, PAWD  comes down. PAWD is always $100\,\%$ or close to $100\,\%$ for PIFS access (not shown).

\subsubsection{Discussion}
From the above discussions on our simulation results, we highlight the following key points. Since keeping PSM to $0\,\%$ is important for the performance of a sensing application, EDCA access is not
a suitable option as it can lead to missed
sensing in almost all cases. So, PIFS-based access should be used for sensing. Very short PAW duration (e.g., 10) is not a good choice, even at a very low sensing load, since it leads to
missed sensing. In fact, with PIFS access, it is better to set the SAW duration to its maximum value 127 to avoid missed sensing. For this SAW duration value, the performance
impact of sensing on data communication in terms of PSO and
throughput is almost identical to smaller SAW durations for which there is no missed sensing.
With PIFS access, when
$\mathit{numApp}$ = 4 and $\mathit{nSTA}$ = 16, the overhead is about $5\,\%$ (see Fig.~\ref{fig:pifs_nsta_pso}) and the
throughput drops by about 8 Mbps (or about $5\,\%$) compared to ``no sensing" case (see Fig.~\ref{fig:pifs_nsta_thrpt}). Considering that this is a very high sensing load situation, the overhead and the throughput drop may be acceptable. Another important thing to note is that the RU size changes at discrete
points (with respect to $\mathit{nSTA}$), and there can be sudden changes in performance or performance may seem counterintuitive at those change points. 
These results also show that a system can be designed with an upper limit on sensing overhead. The AP in such a system would allow
the sensing load to increase (by having more applications, stations, or more antennae) until the sensing overhead limit is reached.


\section{Conclusion}
\label{sec-conclusion}
IEEE 802.11bf is a relatively new standard for Wi-Fi sensing. While integrating sensing with communication in a Wi-Fi network
leads to more efficient use of spectrum and hardware, it also
contributes to communication overhead. Although the \ac{TGbf} has
carefully designed the IEEE 802.11bf protocol to limit the
overhead, there is no sensing performance analysis of the protocol
and its impact on data communication available in the literature. 
Hence, those are the focus of this paper.
Our simulation results show that when NDPA sounding phase with
reporting is enabled, EDCA access is not
suitable for sensing, since it can lead to missed sensing. Also,
a very short SAW duration (e.g., 10), even at a low sensing load,
is not a good choice since it leads to missed sensing. A good
rule of thumb is to have PIFS access with SAW duration set to a large value
(e.g., its maximum value of 127), which ensures $0\,\%$ PSO in almost all cases.

\vspace{-0.1in}
\bibliographystyle{IEEEtran}

\bibliography{master}

\end{document}